\documentstyle[aps,epsf,prd,eqsecnum,multicol]{revtex}
\def\bm{\bbox}

\newcommand{\omegap}{p}
\begin{document}
\thispagestyle{empty}
{\baselineskip-4pt
\font\yitp=cmmib10 scaled\magstep2
\font\elevenmib=cmmib10 scaled\magstep1  \skewchar\elevenmib='177
\leftline{\baselineskip20pt
\hspace{10mm} 
\vbox to0pt
   { {\yitp\hbox{Osaka \hspace{1.5mm} University} }
     {\large\sl\hbox{{Theoretical Astrophysics}} }\vss}}

\rightline{\large\baselineskip20pt\rm\vbox to20pt{
\baselineskip14pt
\hbox{OU-TAP-190}
\hbox{YITP-02-72}
\vspace{1mm}
\hbox{\today}\vss}}%
}
\vskip8mm
\begin{center}{\large \bf Braneworld reheating in the bulk inflaton model} 
\end{center}
\vspace*{4mm}
\centerline{\large 
Yoshiaki Himemoto$^*$
and Takahiro Tanaka$^\dag$
\kern-0.8ex${}^{\star}$}

\vspace*{4mm}
\centerline{\em Department of Earth and Space Science, Graduate 
School of Science}
\centerline{\em Osaka University, Toyonaka 560-0043, Japan} 

\centerline{\em ${}^{\star}$Yukawa Institute for Theoretical Physics,
Kyoto University, Kyoto 606-8502, Japan}

\begin{abstract}
In the context of the braneworld inflation driven by 
a bulk scalar field, we study the energy dissipation 
from the bulk scalar field into the matter on the brane 
in order to understand the reheating after inflation. 
Deriving the late-time behavior of the bulk field with dissipation
by using the Green's function method, we give a rigorous
justification of the statement that 
the standard reheating process is reproduced in 
this bulk inflaton model 
as long as the Hubble parameter on the brane and the mass of the bulk scalar
field are much smaller than the 5-dimensional inverse curvature
scale. Our result supports the idea that 
the brane inflation model caused by a bulk scalar field 
is expected to be a viable 
alternative scenario of the early universe.
\end{abstract}
\begin{multicols}{2}
\section{Introduction}
The braneworld scenario has opened up a new perspective
on higher dimensional theory~\cite{Horava:1996qa}. 
In particular, a model proposed by Randall and
Sundrum (RS2) \cite{Randall:1999vf} has ignited active research 
on braneworld cosmology because of its attractive 
features~\cite{Shiromizu:2000wj,Garriga:2000bq,Garriga:2000yh,cosmology}. 

An alternative scenario of the inflationary
universe in the context of the RS2 model was proposed by Himemoto 
and Sasaki\cite{Himemoto:2001nd}.
In this scenario, inspired by higher dimensional gravitational theory 
including a dilatonic scalar field or higher curvature terms, 
a 5-dimensional Einstein-scalar system was studied.  
Considering a minimally coupled massive scalar field $\phi$
for a 5-dimensional scalar
field as a toy model, 
a solution of the field equation in 
the anti-de Sitter background with an inflating brane
was found under the assumption of a separable form of solution. 
It was shown that a 5-dimensional scalar
field, which we call a bulk scalar field, can realize 
slow-roll inflation as long as $|m^{2}|/H^{2} \ll 1$ is satisfied.
Here, $|m^{2}|$ is the mass squared of the bulk scalar field and $H$ is 
the Hubble parameter on the brane.

As a next step, in order to show the generality of this inflation model,
we investigated the dynamics of a bulk scalar field without 
assuming a separable form of solution,  
focusing on the cases with $|m^{2}|\ell^{2}\ll 1$ and $H^{2}\ell^{2}\ll 1$.  
There, by analyzing the Green's function for the bulk scalar field, 
the late-time behavior of the bulk scalar field was clarified,  
and it was shown that the bulk scalar field effectively 
behaves as a 4-dimensional scalar field $\Phi$ with mass
$m_{\mathrm{eff}}=m/\sqrt{2}$
during both the slow-rolling phase $|m^{2}|/H^{2} \ll 1$ and the rapidly 
oscillating phase $|m^{2}|/H^{2} \gg 1$,   
irrespective of the initial field configuration \cite{Himemoto:2001hu}. 
Moreover, the lowest order back reaction to the geometry starting with the
second order in the amplitude of $\phi$
was found to be consistently represented by a 4-dimensional effective 
theory with the same 4-dimensional 
scalar field $\Phi$ mentioned above. 
The overall normalization of $\Phi$
is related to $\phi$ by a simple scaling $\Phi=\sqrt{\ell} \phi$,
where $\ell$ is the curvature radius of $AdS_5$.
We showed that a bulk scalar field in the braneworld 
can mimic 4-dimensional
inflaton dynamics, at least as long as the Hubble parameter and 
the mass of the bulk scalar field are sufficiently small 
compared with $\ell^{-1}$. 
We also studied the quantum fluctuations
in this inflation model in order to see
whether this model is observationally acceptable or not. 
Then it was shown that the correction due to the 5-dimensional nature
of the inflaton field is small, as long as $H^{2}\ell^{2} \ll 1$ and 
$|m^{2}|\ell^{2} \ll 1$ \cite{Sago:2001gi}.
Kobayashi, Koyama, and Soda also considered the quantum fluctuations
of a massless bulk scalar field in the context of this inflation model
and obtained a consistent result\cite{Kobayashi:2001yh}. 

In \cite{Yokoyama:2001nw,dronhime}, assuming the dominance of a single
oscillation mode, 
it was shown that braneworld reheating also mimics the 
4-dimensional standard process. 
In this short paper, 
in order to give a rigorous justification of the result
 obtained in~\cite{Yokoyama:2001nw,dronhime}, 
we investigate the late-time behavior of the evolution of a bulk 
scalar field with arbitrary, regular initial data using 
the Green's function method.  
We find that the argument of the 
correspondence between the 5-dimensional and 4-dimensional 
systems mentioned in the 
previous paragraph can be extended to include the reheating process.

\end{multicols}
-----------------------------------\hfill
\newline
$^*${E-mail:himemoto@vega.ess.sci.osaka-u.ac.jp}
\hspace{20mm}
$^\dag${E-mail:tanaka@yukawa.kyoto-u.ac.jp}$~~$

\newpage
\begin{multicols}{2}

The organization of this paper is as follows.
In Sec.~II, we review the basic picture of the reheating  scenario
after braneworld 
inflation driven by a bulk scalar field as proposed in \cite{Yokoyama:2001nw}.
In Sec.~III, we consider the late-time behavior of the bulk scalar field
on the brane by using the properties of the retarded Green's function.
Then we explain how to construct the Green's function for 
a scalar field with dissipation.
Subsequently, we discuss an effective description of the dynamics from the
4-dimensional point of view and show that the conventional reheating
process is effectively reproduced in the braneworld. 
Section IV is devoted to a summary 
and discussion.

\section{model of reheating after braneworld inflation}

We briefly review the model 
for the reheating process after braneworld
inflation driven by a bulk scalar field discussed
in Ref.~\cite{Yokoyama:2001nw}. 
We consider a 5-dimensional Einstein-scalar system
in a braneworld scenario of the RS2 type. 
The bulk geometry satisfies the 5-dimensional Einstein equations
\begin{equation}
  R_{ab}-{1\over2}g_{ab}R+\Lambda_5g_{ab}
=\kappa_5^2\left[{T}_{ab}+S_{a b} \delta(r-r_0)\right]\,,
\label{bulkeq}
\end{equation}
where $r$ is the coordinate normal to the brane, and
the brane is assumed to be located at $r=r_0$ at the 
fixed point of $Z_2$ symmetry.
$S_{ab}$ represents the energy-momentum tensor composed of 
the brane tension $\sigma$ and 
the contribution of 
matter fields confined on the brane $\tau_{ab}$:
\begin{equation}
S_{a b} = -\sigma q_{a b}+\tau_{ab}\,,
\label{matter}
\end{equation}
where $q_{ab}$ is the induced metric on the brane.
$T_{ab}$ is the energy-momentum tensor 
of the bulk scalar field. 
For simplicity, we assume a minimally coupled massive scalar field 
with an interaction on the brane. 
We set 
$\Lambda_5=-\kappa_5^4\sigma^2 / 6$
so as to recover the Randall-Sundrum flat braneworld
when $T_{ab}$ and $\tau_{ab}$ vanish.

In order to represent the energy dissipation from the bulk scalar field
$\phi$ to the matter fields on the brane, 
the equation for $\phi$ is modified to 
\begin{equation}
 (\Box_5-m^{2})\phi
=\Gamma_d \,\ell \,  
\delta \left(r-r_{0}\right)\dot{\phi}\,,
\label{kgv}
\end{equation}
where $\ell=6/(\kappa_5^2\sigma)$ is the $AdS_5$
curvature radius, and $m$ is the mass of the bulk scalar field.
An overdot indicates differentiation with respect to the cosmological 
proper time. 
$\Gamma_d$ represents the decay constant. 
Integrating Eq.~(\ref{kgv}) together with $Z_2$ symmetry,
we obtain the boundary condition 
\begin{equation}
\partial_{r} \phi|_{r=r_{0}}
=\frac{\Gamma_d \ell}{2}\dot{\phi}(t,r_0).
\label{phiw}
\end{equation}
In the analyses below, we neglect the terms quadratic in $\Gamma_d$. 
Here, we do not consider the energy release to the
bulk space, although such a dissipation 
process in the bulk may be worthy of investigation.

\section{Bulk scalar dynamics with dissipation}
We consider solutions of Eq.~(\ref{kgv})
under the conditions 
\begin{equation}
m \ell\,, 
H\ell\,,~ {\rm and}~ \Gamma_{d} \ell \ll 1\,, 
\label{condition}
\end{equation}
where $m$ and $H$ are, respectively, 
the mass of the bulk scalar field and 
the expansion rate of the 4-dimensional metric induced on the brane. 
Then we investigate the late-time behavior of solutions 
by extending the Green's function method used 
in Ref.~\cite{Himemoto:2001hu}. 

%


\subsection{Initial value problem}
We consider the field equation
\begin{equation}
 {\cal L}_X\, \Psi(X)\equiv
  \left[-\partial_t^2+\partial_y^2-W(y)-U(y)\partial_t\right]\Psi(X) 
  =0\, ,   
\label{fieldeq}
\end{equation}
where $X$ denotes $\left\{ t,y \right\}$.
We assume that $W(y)$ and $U(y)$ are functions whose 
asymptotic value at $|y|\to \infty$ is supposed to be a constant. 
We solve this field equation by imposing
$Z_2$ symmetry at $y=0$. 

Let us consider the retarded Green's function
that satisfies 
\begin{equation}
  {\cal L}_X G(X,X')=  -\delta(t-t')\delta(y-y'),
\label{Greeneq}
\end{equation}
with the causal condition that $G(X,X')=0$ for $X'$ not in the
causal past of $X$.
Note that this Green's function satisfies reciprocity
for the time reversal operation $t\to -t$, namely, $G(t,y,t',y')=G(-t',y',-t,y)$. 
Using this reciprocity, 
the evolution of $\Psi$ for any given initial data 
can be described as
\begin{eqnarray}
 \Psi  &= &  
   \int_{t'=t_i} \Biggl[  G(X,X'){\partial \Psi(X')\over \partial t'} 
      -{\partial G(X,X')\over \partial t'}\Psi(X') \cr
  &&\hspace*{2cm}
      +U(y') G(X,X')\Psi(X') \Biggr] dy',
\end{eqnarray}
where $t_i$ is the initial time.
This representation of $\Psi$ implies
that the asymptotic behavior of $\Psi$ can be understood 
by the late-time behavior of $G(X,X')$. 

Let us now construct the Green's function that satisfies Eq.~(\ref{Greeneq}).
Because of time translation invariance, 
we can express the Green's function 
in terms of Fourier transformation as 
\begin{equation}
  G(X,X')={1\over 2\pi}\int_{\cal{C}} d\omegap\, G_\omegap(y,y') 
       e^{-i\omegap(t-t')}\,, 
\label{formalG}
\end{equation}
where we choose the path ${\cal{C}}$ so that the retarded condition 
is satisfied; 
namely, ${\cal{C}}$ 
runs from $\omegap=-\infty$ to 
$\omegap=+\infty$ on the complex plane for the integrand $ G_\omegap$
to contain neither pole nor the branch cut above ${\cal{C}}$. 

The equation for $G_{\omegap}$ follows from Eqs.~(\ref{Greeneq}) and  
(\ref{formalG}) as
\begin{eqnarray}
 &&{\cal L}^{(\omegap)}_y G_{\omegap}(y,y') \cr
 &&\quad \equiv 
 [\partial_y^2+\omegap^2-W(y)+i\omegap U(y)]G_{\omegap}(y,y')\cr
     &&\quad =-\delta(y-y').   
\end{eqnarray}
The homogeneous equation ${\cal L}^{(\omegap)}_y u(y)=0$ has  
two independent solutions which respectively behave
asymptotically as 
$  u^{({\mathrm{out}})}_{\omegap} (y)  \sim  e^{ik|y|}$ and  
$u^{({\mathrm{in}})}_{\omegap} (y)  \sim  e^{-ik|y|}$ at 
$|y|\to \infty$, where we defined $k^2=
\displaystyle\lim_{|y|\to\infty}\omegap^2-W(y)+i\omegap U(y)$. 
We can describe the solution satisfying the condition of $Z_2$ symmetry as 
a linear combination of these two independent solutions, 
$ u_{\omegap}^{(Z_2)}(y)= u^{({\mathrm{out}})}_{\omegap} (y) 
-\gamma_{\omegap} u^{({\mathrm{in}})}_{\omegap}(y)$\,,  
where
$\gamma_\omegap$ is determined by the $Z_2$ symmetry condition 
\begin{equation}
{\cal D}[u_{\omegap}^{(Z_2)}(y)]
\equiv [\partial_y - \hat W +i\omegap\hat U] 
u^{(Z_2)}_{\omegap}(y)|_{y=0+}=0\,,
\label{poleEq}
\end{equation}
where $\hat W$ and $\hat U$ are the coefficients 
in front of $\delta(y)$  
contained in $W(y)/2$ and $U(y)/2$, respectively. 
With these mode functions, we can express 
$G_{\omegap}(y,y')$ satisfying the boundary condition
at infinity and on the brane as
\begin{eqnarray}
  G_\omegap(y,y') &= & {1\over -2i\omegap\gamma_\omegap}
      \biggl[u_\omegap^{({\rm{out}})}(y) u^{(Z_2)}_\omegap(y')
          \theta(|y|-|y'|)\cr
&&\hspace*{1cm}
       +u^{(Z_2)}_\omegap(y) u_\omegap^{({\rm{out}})}(y')
          \theta(| y'|-|y|)\biggr]
\end{eqnarray}
with
$ \gamma_{\omegap}={\cal D}[u_{\omegap}^{({\rm{out}})}(y)]
         / {\cal D}[u_{\omegap}^{({\rm{in}})}(y)]$.

The late-time behavior of $\Psi$ is understood by 
investigating the structure of
singularities such as poles and branch cuts in $G_\omegap (y,y')$. 
The singularity
on the complex $\omegap$ plane with the largest imaginary part dominates
the late-time behavior. 
In particular, poles exist 
at the values of $\omegap$ for which 
\begin{equation}
 {\cal D}[u_{\omegap}^{({\rm{out}})}(y)]=0
\label{setup}
\end{equation}
holds.

\subsection{Application to braneworld}
Using the formulas developed in the previous section, 
we consider the evolution of a massive bulk scalar field with
dissipation, which well approximates 
the situation where $\phi$ oscillates at the bottom of the
potential. 
We assume that the background spacetime is 
5-dimensional anti-de Sitter space 
with a boundary de Sitter brane given by 
$ds^{2}=dr^{2}+\{H\ell \sinh(r/\ell)\}^2
(-dt^{2}+H^{-2}e^{2Ht}d{\bm{x}}^{2}_{(3)})
$~\cite{Garriga:2000bq}. 
In order to use the formulas presented 
in the preceding subsection, we introduce the conformal 
coordinates $\{\tau,y\}$ defined by 
$\tau \equiv Ht$ and 
$R(y)\equiv \ell\sinh^{-1}(|y|+y_0) = \ell \sinh (r/\ell)$, 
where $y_0$ is specified by 
$\sinh(y_0)=\sinh^{-1}(r_0/\ell)=H\ell$.
Then the metric becomes 
\begin{equation}
ds^{2}=R(y)^{2}
\left(dy^{2}-d\tau^{2}+e^{2\tau}\,d\mbox{\boldmath
$x$}_{(3)}^{2}\right)
\,. 
\end{equation}

Setting $\phi=R^{-3/2}e^{-3\tau/2}\Psi(y,\tau)$, Eq.~(\ref{kgv}) 
reduces to Eq.~(\ref{fieldeq}) with 
\begin{eqnarray}
{W(y)}&=&{{15+4m^{2}\ell
^{2}}\over 4 \sinh^{2}(|y|+y_0)}\cr
 &&\hspace{1cm}-
\left( {3\sqrt{1+H^{2}\ell^{2}} \over H\ell}
+{3\over 2}\Gamma_d \ell \right)\delta(y),\\ 
U(y)&=& \Gamma_{d} \ell \, \delta\left(y \right)\,.
\end{eqnarray}
We note  that 
both ${W}(y)$ and ${U}(y)$ vanish for $|y|\rightarrow
\infty$, and hence the conditions that 
they asymptotically become constant are satisfied. 
Although $U(y)$ is localized on the brane in the present model, 
dissipation in the bulk can also be discussed by using the same 
technique as long as $U(y)$ 
satisfies this property. 

The mode solution satisfying the outgoing asymptotic behavior is 
\begin{eqnarray} 
 u_\omegap^{({\mathrm{out}})}(y) 
& = & \Gamma(1-i\omegap)P_{\nu-1/2}^{i\omegap}[\coth(|y|+y_0)]
\,,
\label{outgoing}
\end{eqnarray}
where 
$\nu=\sqrt{m^2\ell^2+4}$ \cite{bateman}.
In the present case, the operator defined in Eq.(\ref{poleEq})
reduces to  
\begin{equation}
 {\cal D}=\partial_y +{3 \over 2}{\sqrt{1+H^{2}\ell^{2}} \over H\ell}
+\left(i\omegap+{3\over2}\right){\Gamma_d \ell\over 2}\,.
\label{operator}
\end{equation}
Then,
from Eqs.~(\ref{setup}) and (\ref{outgoing})
under the condition (\ref{condition}), 
we obtain the location of the poles with the largest 
imaginary parts in the complex $\omegap$ plane as 
\begin{equation}
 \omegap_{\pm} \approx {1\over H}\left[
  -i {\Gamma_d \over 2}\pm\sqrt{{m_{\mathrm{eff}}^2-{(3H+\Gamma_d)^2\over 4}}}
 \right]\,,
\label{ppm}
\end{equation}
where
\begin{equation}
m_{\mathrm{eff}}^{2}
={m^{2}\over 2}.  
\label{mass}
\end{equation}
Here the terms neglected are quadratic or higher order in $\ell$. 
In addition to two poles, 
there is an infinite sequence of poles. 
However, 
they do not give a dominant contribution to
the late-time behavior since
they have smaller imaginary parts. (See Ref.~\cite{Himemoto:2001hu}.)
Therefore the asymptotic behavior of the field $\phi$ 
is dominated by the poles $\omegap_\pm$, and hence we have 
\begin{equation}
\phi\propto 
e^{-{3\over2}\tau } G(X,X') \propto 
   e^{-({3\over 2}+i\omegap_{\pm})H t}. 
\label{asym2}
\end{equation}

\subsection{Effective 4-dimensional description}
In the preceding section, we confirmed that the late-time 
behavior of a bulk scalar field on the brane can be 
effectively described by
a 4-dimensional scalar field $\Phi$ which satisfies
\begin{eqnarray}
\ddot\Phi+(3H+\Gamma_d)\dot\Phi+m_{\rm{eff}}^{2}\Phi
=0.
\label{infdynamics}
\end{eqnarray}
Our main interest here is to compare 
the law of cosmic expansion on the brane 
to that expected in the usual 4-dimensional model
with the scalar field $\Phi$. 
First we present a rather general framework 
to discuss the cosmic expansion of a homogeneous universe 
realized on the brane without specifying the explicit form of 
the energy-momentum tensor in the bulk. 

We introduce a unit vector in the time direction $u^a$ 
parallel to the brane. Toward the outside of the brane 
$u^a$ is extended so as to satisfy $u^a{}_{;b}n^b=0$, 
where $n_{b}$ is the outgoing unit normal vector of the brane.  
Then, considering integration of the 5-dimensional conservation law 
\begin{equation}
 0=\int \sqrt{-g}\, d^5x\, [S_a{}^b\delta(r-r_0)
   +{T}_a{}^b]_{;b} u^a 
\end{equation}
for a thin region surrounding the brane, we obtain 
\begin{equation} 
  \dot \rho+4H\rho+H \tau^{\mu}_{~\mu}= 
  2 {T}_{ab} u^a n^b\,, 
\label{cons4d}
\end{equation}
where $\rho:=u^\mu u^\nu \tau_{\mu\nu}$, and latin indices 
run over 4-dimensional coordinates on the brane.  

The conservation law for the matter field localized on the 
brane $\tau^{\mu\nu}{}_{;\nu}= 0$ is violated  
because there is energy 
transfer between the brane and bulk fields.
However, from the Bianchi identity 
the 4-dimensional conservation law is guaranteed to be preserved 
in total. 
The 4-dimensional effective Einstein equation is derived from 
Eq.(\ref{bulkeq}) 
as\cite{Shiromizu:2000wj} 
\begin{equation}
  {}^{(4)}G_{\mu\nu}=
\kappa_4^2(\tau_{\mu\nu}
           +\tau_{\mu\nu}^{(\pi)}+\tau^{(s)}_{\mu\nu}+\tau_{\mu\nu}^{(E)}
            ),  
\end{equation}
where $\kappa_{4}^{2} = {\kappa_5^2/\ell}$ 
and 
\begin{eqnarray}
\tau_{\mu \nu}^{(\pi)} &=& 
 -{\kappa_5^2 \ell\over 24}\bigl[6\tau_{\mu \alpha}\tau_{\nu}^{\ \alpha}
 -2\tau \tau_{\mu \nu} 
 -q_{\mu \nu}(3\tau_{\alpha \beta}\tau^{\alpha\beta}
  -\tau^{2})\bigr],\hspace{-1cm}
\nonumber\\
  \tau_{\mu\nu}^{(s)} &= &{2\ell\over 3}
   \left[{T}_{ab} q^a{}_\mu q^b{}_\nu
          +q_{\mu\nu}\left({T}_{ab} n^a n^b
            -{1\over 4} {T}^{a}{}_a
             \right) \right],\hspace{-0cm}
\nonumber\\
\tau_{\mu \nu}^{(E)}&=& \,-{\ell \over \kappa_5^2} {}^{(5)}C_{r b r d}
\,q_{\mu}^{b}\,q_{\nu}^{d}\,.
\end{eqnarray}
Here ${}^{(5)}C_{r b r d}$ is the 5-dimensional Weyl tensor with its
two indices projected in the normal direction.
Then we have 
\begin{equation}
 \tau^{({\rm{tot}})\mu\nu}{}_{|\nu}=0
\label{Excons}
\end{equation}
with 
\begin{equation}
 \tau^{({\rm{tot}})}_{\mu\nu}:=\tau_{\mu\nu}
        +\tau^{(\pi)}_{\mu\nu}+\tau^{(s)}_{\mu\nu}+\tau_{\mu\nu}^{(E)}\,.  
\end{equation}
Here the vertical bar 
represents covariant differentiation with respect 
to the 4-dimensional induced metric $q_{\mu\nu}$. 
In a homogeneous universe, 
the temporal component of the conservation law reduces to 
\begin{equation}
 \dot\rho^{({\rm{tot}})}+ 4H\rho^{({\rm{tot}})}+H \tau^{({\rm{tot}})\mu}{}_{\mu}=0,
\end{equation}
where $\rho^{({\rm{tot}})}:=u^\mu u^\nu \tau^{({\rm{tot}})}_{\mu\nu}$. 
Similarly, we define the energy density for each component by 
$\rho^{(i)}:=u^\mu u^\nu \tau^{(i)}_{\mu\nu}$.  
Then, using the facts that
$  \rho^{(\pi)}=  \rho^2/2\sigma$,
$\tau^{(\pi)\mu}{}_{\mu}= 
{(\rho/ \sigma)} (\tau^{\mu}{}_{\mu}+2\rho) 
$~\cite{Shiromizu:2000wj},  
and 
$
 \tau^{(s)\mu}{}_{\mu} = 2 \ell {T}_{a b} n^a n^b 
$, 
we find that 
the effective energy density of the bulk field 
$\rho_{\rm{eff}}:=\rho^{(E)}+\rho^{(s)}$ satisfies  
\begin{equation}
 \dot \rho_{\rm{eff}}+4H\rho_{\rm{eff}}
  =-2\left(1+{\rho\over\sigma}\right){T}_{ab} u^a n^b  
  -2 H\ell {T}_{a b} n^a n^b.
\label{cons5d}
\end{equation}

Now let us return to our specific model, in which
the bulk energy-momentum tensor is given by 
\begin{eqnarray}
T_{a b} = \phi_{,a} \phi_{,b}- g_{a b}\left( {1\over2}
g^{c d}\phi_{,c} \phi_{,d}+ {1\over2}m^{2}\phi^{2}\right). 
\end{eqnarray}
Then we find 
\begin{eqnarray}
T_{ab} u^a n^b & = & 
   {\Gamma_d \ell\over 2}\dot\phi^2,\cr
T_{ab} n^a n^b & = & {1\over 2}\dot\phi^2-{1\over2}m^{2}\phi^{2}, 
\end{eqnarray}
where the terms quadratic in $\Gamma_d \ell$ were neglected. 
If we set the identification~\cite{Himemoto:2001hu}
\begin{equation}
 \Phi=\sqrt{\ell}\,\phi\,,
\end{equation}
from Eqs.~(\ref{cons4d}) and (\ref{cons5d}), we obtain 
the formulas for the cosmic expansion 
valid in the low energy regime (\ref{condition}) as 
\begin{eqnarray}
H^2& = & 
{{\kappa _4^2} \over 3}\left( \rho_{\rm{eff}}
+\rho \right), \nonumber\\
\dot \rho_{\rm{eff}} +4H \rho_{\rm{eff}}
&=&-H\left(\dot\Phi^2-4V_{\mathrm{eff}}\right) 
     -\Gamma_d  \dot \Phi ^2,\nonumber \\
\dot\rho+4H\rho+H \tau^{\mu}_{~\mu}
&=& \Gamma_d \dot \Phi^2,
\end{eqnarray}
where
$V_{\mathrm{eff}}(\Phi)={1\over2} m_{{\rm{eff}}}^{2} \Phi^{2} 
=(\ell/4)m^2\phi^2$. 
This set of equations is 
the same as that satisfied by a 4-dimensional model
having a scalar field $\Phi$ with mass squared $m^2_{\mathrm{eff}}=m^2/2$ and
decay width $\Gamma_d/2$. 
To conclude, provided that the conditions (\ref{condition}) 
are satisfied,
the effective dynamics of the Einstein-scalar system
on the brane is indistinguishable from
a 4-dimensional theory 
even if we consider energy dissipation from 
the bulk scalar field to the matter fields on the brane. 

\section{Summary and Discussion}
We have investigated the late-time behavior of a bulk scalar field
with dissipation to the matter fields on the brane. 
We have shown that a bulk scalar field observed on the brane
effectively behaves as a 4-dimensional scalar field. 
Furthermore, we have shown that the set of 
equations to determine the cosmic expansion 
is also indistinguishable from that of the corresponding 
standard 4-dimensional model. 
This result reinforces the speculation previously presented 
in Refs.~\cite{Yokoyama:2001nw,dronhime}.
Alhough we have analyzed perturbatively only 
the dynamics of a bulk scalar field on a fixed 
5-dimensional anti-de Sitter background with 
a boundary de Sitter brane, 
our result suggests that 4-dimensional 
inflaton dynamics including the reheating era
is effectively reproduced by the dynamics of a 
5-dimensional scalar field. 
Thus the brane inflation model caused by a bulk scalar field 
is expected to be a viable alternative scenario of the early universe.

Genuine braneworld effects 
that can be used to test the scenario are, however, 
in the deviations from the standard model.  
Our analysis shows that the corrections 
are suppressed by a factor $H^2\ell^2$ or $|m^2|\ell^2$.  
We did not go into detail about this correction in 
this paper. Here we just mention that 
we have to consider the dynamics of a bulk scalar field 
with a general Friedmann-Robertson-Walker(FRW) brane boundary 
when we discuss corrections of this order, 
because the expected suppression of the correction 
due to the variation of the expansion rate 
is of the same order: ${\dot{H}\ell^{2}}\sim H^2\ell^2$.  

Another regime 
where $H^{2}\ell^{2}\gg 1$ and $|m^{2}|\ell^{2}\gg 1$ 
in this bulk inflaton model is also interesting.
Even in this regime, slow-roll inflation seems to be realized on the brane 
as long as $|m^{2}|/H^{2} \ll 1$ is satisfied~\cite{Himemoto:2001nd}, 
although many unsolved issues remain there.  

Last we should mention dissipation in the bulk.
When we consider a specific well-motivated model, the bulk inflaton field may
naturally have interaction with the other fields in the bulk. 
Once a model is specified, we would be able to estimate the 
strength of the coupling.  
Thus, investigating dissipation in the bulk might give an important 
constraint on the construction of a realistic model. 
The formulas developed in this paper might also be useful 
for this purpose. 
Investigations in this direction are in progress.

\section*{Acknowledgments} 
We are grateful to M.Sasaki and J.Yokoyama for useful communications.
T.T. would like to thank L.~Sorbo and D.~Langlois for 
useful discussions during his stay at IAP. 
To complete this work, the discussion during
the YITP workshop YITP-W-01-15 on ``Braneworld - Dynamics of 
spacetime boundary'' was useful. 
This work was supported in part 
by the Monbukagakusho Grant-in-Aid 
No.~14740165(T.T.).

\end{multicols}
\end{document}